\documentclass[aps,pre,twocolumn,showpacs,showkeys]{revtex4}

\usepackage{graphicx}
\usepackage{bm}
\usepackage{amsmath}
\usepackage{dcolumn}

\begin{document}

\title{Observation of the Nuclear Magnetic Octupole Moment of $^{87}$Rb
from Spectroscopic Measurements of Hyperfine Intervals}

\author{Vladislav Gerginov}\email[email:]{vgergino@gmail.com}
\author{Carol E. Tanner} \email[email:]{carol.e.tanner.1@nd.edu}
\author{W.R.Johnson} \email[email:]{johnson@nd.edu}
\affiliation {Department of Physics, 225 Nieuwland Science Hall,
University of Notre Dame, Notre Dame IN 46556}

\begin{abstract}
The magnetic octupole moment of $^{87}$Rb is determined from hyperfine
intervals in the $5p\ ^2\! P_{3/2}$ state measured by Ye {et al.} [Opt.\ Lett. {\bf 21}, 1280 (1996)].
Hyperfine constants $A = 84.7189(22)$~MHz, $B = 12.4942(43)$~MHz, and $C = -0.12(09)$~kHz are
obtained from the published measurements.
The existence of a significant value for $C$ indicates the presence of a nuclear
magnetic octupole moment $\Omega$. Combining the hyperfine constants
with atomic structure calculations, we obtain $\Omega = -0.58(39)$~b$\mu_N$.
Second-order corrections arising from interaction with the nearby
$5p\ ^2\! P_{1/2}$ state are found to be insignificant.
\end{abstract}


\maketitle

\section{Introduction}
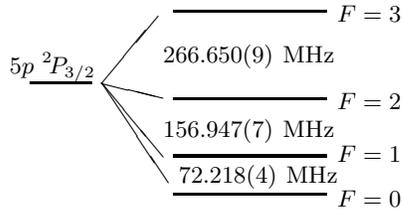
\begin{figure}
\setlength{\unitlength}{1ex}
\begin{picture}(40,22)
\thicklines
\put(0,14){$5p\ ^2\!P_{3/2}$}
\put(2,13){\line(1,0){6}}
\put(32,19){\small $F=3$}
\put(15,15){\small 266.650(9) MHz}
\put(16,20){\line(1,0){15}}
\put(32,10.5){\small $F=2$}
\put(15,7.7){\small 156.947(7) MHz}
\put(16,11.5){\line(1,0){15}}
\put(32,5.3){\small $F=1$}
\put(16.5,3.3){\small 72.218(4) MHz}
\put(16,5.9){\line(1,0){15}}
\put(32,0.9){\small $F=0$}
\put(16,2.2){\line(1,0){15}}
\thinlines
\put(9,13){\line(5,6){5.5}}
\put(9,13){\line(2,-3){6.5}}
\put(9,13){\line(5,-6){6}}
\put(9,13){\line(4,-1){6}}
\end{picture}
\caption{Hyperfine structure of the $5p\ ^2\! P_{3/2}$ state in $^{87}$Rb from
\citet{YSJH:96}.\label{fig1}}
\end{figure}

We report values of the nuclear  magnetic octupole moment of $^{87}$Rb
determined from measured hyperfine intervals. During an earlier investigation
of the hyperfine structure of atomic $^{133}$Cs \cite{GDT:03}, we
discovered existing measurements of the hyperfine structure of the $5p\ ^2\! P_{3/2}$ state
of $^{87}$Rb ($I = 3/2$, $F = 0,\, 1,\, 2,\, 3$) by \citet{YSJH:96} with a frequency
resolution of several kHz.
The measured hyperfine intervals $\Delta W_{F} = W_F - W_{F-1}$ are shown
in Fig.~\ref{fig1}.
These intervals were decomposed in terms of nuclear magnetic dipole
and electric quadrupole coupling coefficients ($A$ and $B$) in Ref.~\cite{YSJH:96}
and the resulting values for $A$ and $B$ are given in the first row of Table \ref{tab1}.
However,
for an atomic state with angular momentum $J \ge 3/2$, there also exists a coupling between
the nuclear magnetic octupole moment and the electronic state, provided $I\ge 3/2$.
Precision measurements of the three hyperfine intervals in $^{87}$Rb are sufficient to
determine three hyperfine constants $A$, $B$ and $C$,
representing the interaction of the nuclear magnetic dipole moment $\mu_I$, electric quadrupole moment $Q$,
and magnetic octupole moment $\Omega$, respectively, with the atomic electrons.
We combine the coefficient $C$,
determined herein, with atomic
structure calculations for neutral $^{87}$Rb to determine the nuclear octupole moment.
The purpose of the present paper is to determine the previously unknown nuclear magnetic octupole
moment of $^{87}$Rb, not to discuss the experimental work by \citeauthor{YSJH:96}.

The (unperturbed) hyperfine structure intervals of an atom
having a single electron outside closed shells and $I=J=3/2$ is given in terms of the hyperfine
constants $A,\, B$ and $C$ by \cite{A:71}
\begin{align}
\Delta W_{3} &=\ 3A + B + 8C   \label{one}\\
\Delta W_{2} &=\ 2A - B - 28C \label{two}\\
\Delta W_{1} &=\ A - B + 56C . \label{three}
\end{align}
Inverting these equations leads to the values of $A,\, B$ and $C$ given in the second row of Table~\ref{tab1}.

\section{Semi-empirical Analysis}

With the aid of approximations developed by \citet{Cas:63},  semi-empirical estimates for the
hyperfine coupling constants can be expressed in terms of the nuclear moments $\mu_I$, $Q$, and
 $\Omega$ of $^{87}$Rb. To evaluate
the hyperfine constants empirically, we start from the nonrelativistic expressions for $A,\, B$ and $C$
\cite[][Chap.\ VIII]{A:71}, which can be written
\begin{align}
A(5p_{3/2}) & =\ \mu_I(\mu_N) \, \frac{16}{45} \left< \frac{1}{r^3} \right>_{5p}\!\! F \times  95.4107\ \text{MHz} \label{aeq} \\
B(5p_{3/2}) & =\  Q(b)\,   \frac{2}{5} \left< \frac{1}{r^3} \right>_{5p}\!\! R \times  234.965\ \text{MHz} \label{beq} \\
C(5p_{3/2}) & =\ \Omega(\mu_N b)\,  \frac{2}{35} \left[\frac{R^2_{5p}(r)}{r^4}\right]_{0}\!\! T \times  3.40718\  \text{Hz}. \label{ceq}
\end{align}
The atomic matrix elements in Eqs.~(\ref{aeq},\ref{beq}) are expressed in atomic units.
In Eq.(\ref{ceq}), $R_{5p}(r)$
is the radial wave function of the $5p$ state and, again, atomic quantities are expressed in atomic units.
The factors F, R, T in the above equations account for relativistic effects.
With aid of the \citeauthor{Cas:63} approximations,
we write
\begin{align}
 \left< \frac{1}{r^3} \right>_{np} & \approx\   \frac{Z_i^2 Z}{3 (n^\ast)^3}\left(1-\frac{d\sigma}{dn}\right)
= 0.95983\, a_0^{-3} \\
  \left[\frac{R^2_{np}(r)}{r^4}\right]_0 & \approx\
 \frac{4 Z_i^2 Z^3}{9  (n^\ast)^3}\left(1-\frac{d\sigma}{dn}\right) = 1393.68\, a_0^{-5} .
\end{align}
In the above expressions, $Z = 37-4 = 33$ is the effective nuclear charge  \cite[][p.\ 166]{A:71},
$Z_i = 1$ is the ionic charge, $n^\ast$
is the effective quantum number of the $5p_{3/2}$ state, and $\sigma$ is the quantum defect.
From \citet{V:97}, we find $n^\ast = 2.293$ and $d\sigma/dn = -0.052$, for the $5p_{3/2}$ state
of Rb.
Using the values $\mu_I = 2.7516 \mu_N$ \cite{R:89} and $Q = 0.1335\ b$ \cite{P:01} for $^{87}$Rb, we obtain
\begin{align}
A(5p_{3/2})  & \approx  89.6\, F \ \text{MHz} \\
B(5p_{3/2})  & \approx  12.0\, R \ \text{MHz} \\
C(5p_{3/2})  & \approx  0.271\, T\ \Omega\ \text{kHz} .
\end{align}
 We estimate the relativistic correction factors $F,\, R,\, T$ by comparing
relativistic and nonrelativistic Hartree-Fock (HF) calculations of
$A,\, B$ and $C$. In this way, we obtain $F = 1.02$,
$R = 1.04$, and  $T=1.08$.  The semi-empirical values
listed in Table~\ref{tab1} include these relativistic corrections.
As seen from the table, the semi-empirical value for $A$ differs
from experiment by 8\%, while the value of $B$ differs by less than
0.1\%.

\begin{table}
\caption{Hyperfine coupling constants $A$, $B$ and $C$ from measured hyperfine intervals of $^{87}$Rb
$5p\ ^2\! P_{3/2}$. First row: results obtained by \citet{YSJH:96}, second row: present analysis, third
row: semi-empirical estimates, fourth row: MBPT results. In the semi-empirical and MBPT calculations,
we use $\mu_I = 2.7516\mu_N$ from \citet{R:89} and $Q = 133.5(5)$~mb from \citet{P:01}.  $\Omega$
is expressed in $\mu_N$b.
\label{tab1}}
\begin{ruledtabular}
\begin{tabular}{cccc}
                     &  $A$ (MHz)   &  $B$ (MHz) & $C$ (kHz)\\
                     \hline
Ref.~\cite{YSJH:96}  &  84.7185(20) & 12.4965(37)&          \\
Present  Work        &  84.7189(22) & 12.4942(43)& -0.12(09)\\
Semi-empirical       &  91.4        & 12.5       &  0.293 $\Omega$\\
MBPT                 &  83.0        & 12.6       &  0.206 $\Omega$
\end{tabular}
\end{ruledtabular}
\end{table}

\section{Relativistic Many-Body Analysis}

In Ref.~\cite{SJD:99}, a relativistic many-body method, referred to as the SD approximation,
in which single and double excitations of  Dirac-Hartree-Fock wave functions are
iterated to all orders in perturbation theory,
was used to predict magnetic-dipole hyperfine constants for low-lying states of alkali-metal atoms
to within a few percent.
The SD value for the magnetic-dipole hyperfine constant of the $5p_{3/2}$
state of $^{85}$Rb from \cite{SJD:99} is $A = 24.5$~MHz.
This value is scaled by the ratio
\[
\frac{g_I[^{87}\text{Rb}]}{g_I[^{85}\text{Rb}]} = \frac{2.7516/(3/2)}{1.3534/(5/2)} = 3.3885
\]
giving $A = 83.0$~MHz for the $5p_{3/2}$ state of $^{87}$Rb shown in the fourth row of Table~\ref{tab1}.
The all-order method described in \cite{SJD:99} was used here to obtain
$B/Q = 94.16$~MHz and $C/\Omega = 0.206$~kHz,
leading to the values of $B$ and $C$ listed in the fourth row of Table~\ref{tab1}.
The theoretical uncertainty in these values is estimated to be 2\%.

Experimental values of the nuclear moments are listed in the first row of Table~\ref{tab2}.
By comparing the all-order SD values of $A/\mu_I$ and $B/Q$  with the experimental values
of $A,\ B$ and $C$ shown in the second-row of Table~\ref{tab1},
we obtain the MBPT values of  nuclear moments $\mu_I$, $Q$ given in the second row
of Table~\ref{tab2}.
The values of $\mu_I$ and $Q$ determined in this way agree to within 2\% with precise measurements.
Comparing the theoretical value of $C/\Omega$ with the experimental value of $C$ given in the second row
of Table~\ref{tab1} leads
to the principal prediction of the present paper: $\Omega$ = -0.58(39) $\mu_N$b.

\section{Estimating Nuclear Moments}
It is of interest to compare values of nuclear moments inferred from
atomic structure calculations with values obtained directly
from nuclear shell-model calculations.
In the extreme shell model \cite{MJ:55}, properties
of $^{87}$Rb can be described assuming a single valence nucleon moving
around an inert core.
According to \citet{S:55}, the shell-model predictions for the nuclear moments
$\mu_I$, $Q$ and $\Omega$ are:
\begin{multline}
\mu_I =  \mu_N  I \times \\
\left\{ \begin{array}{ll}
  g_L +  (g_S-g_L)/(2I),    & I=L+1/2\\
  g_L -  (g_S-g_lL)/(2I+1), & I=L-1/2
\end{array}
\right.  \label{aaa}
\end{multline}
\begin{equation}
Q = - \frac{2I-1}{2I+2} g_L  \langle r^2 \rangle\label{bbb}
\end{equation}
\begin{multline}
\Omega = \frac{2}{3} \mu_N \frac{(2I-1)}{(2I+4)}\langle r^2 \rangle \\
\times
\left\{ \begin{array}{ll}
(I+2)[ (I-3/2)g_L+g_S], & I=L+1/2 \\
(I-1)[ (I+5/2)g_L-g_S], & I=L-1/2
\end{array} \right.  .\label{ccc}
\end{multline}
For $^{87}$Rb, the unpaired $p_{3/2}$ proton has total angular momentum $I = 3/2$,
orbital angular momentum $L=1$, and spin angular momentum $S=1/2$.
The proton spin gyromagnetic ratio is $g_S = 5.585694701(56)$ \cite{MTN:08}
and the orbital gyromagnetic ratio is $g_L = 1$.
Using the value of the mean-squared nuclear radius $\langle r^2 \rangle = 0.180$~b from \cite{BMPS:98},
we obtain the values of the three nuclear moments given in the third row of Table~\ref{tab2}.
Comparing the shell-model value of $Q$ with the experimental value shown in Table~\ref{tab2}, it is clear
that nuclear many-body effects modify both sign and magnitude of the shell-model prediction.
Therefore, the difference in sign and magnitude
between the value of $\Omega$ predicted here and given in the second row of Table~\ref{tab2}
and the shell-model value
shown in the third row is not particularly surprising!

\begin{table}
\caption{Nuclear moments of $^{87}$Rb. First row: mean experimental values,
second row: values calculated using MBPT matrix elements and the hyperfine
constants given in Table~\ref{tab1},
third row: values calculated in the extreme nuclear shell model Eqs.(\ref{aaa}-\ref{ccc}).
  \label{tab2}}
\begin{ruledtabular}
\begin{tabular}{cccc}
           & $\mu_I$ ($\mu_N$) &  $Q$ (mb) & $\Omega$ ($\mu_N$b) \\
\hline
Expt.      & 2.751639(2)       &  133.5(5)  &      -   \\
MBPT       & 2.70              &  135       & -0.58(39)\\
Shell-model&  3.79             &  -72       & 0.30     \\
\end{tabular}
\end{ruledtabular}
\end{table}

\section{Second-Order Corrections}
Second-order hyperfine corrections will modify Eqs.~(\ref{one}-\ref{three}) and possibly influence
the value of $C$ extracted from these equations. In this section, we investigate the influence of
second-order corrections and show that they have a negligible effect on the $5p_{3/2}$ hyperfine
constants in $^{87}$Rb.
The second-order correction to the energy of a state $|0 \rangle$ can be written
\begin{equation}
W^{(2)} = \sum_n \frac{\langle 0 |H_\text{hf}| n\rangle \langle n | H_\text{hf} | 0\rangle}
{E_0-E_n} ,
\end{equation}
where $H_\text{hf}$ is the hyperfine interaction Hamiltonian.
For the case of interest here, $|0\rangle$ is a particular hyperfine substate of the $5p_{3/2}$ state
and the sum over states $|n\rangle$ is restricted to substates of the nearby $5p_{1/2}$ state.
Only the two substates $F=1$ and $F=2$ of the $5p_{3/2}$ states are modified by interaction
with the $5p_{1/2}$ state. Following the discussion in \cite{BDJ:08}, we find
\begin{equation}
 W_F^{(2)} = \begin{cases}
 \frac{\textstyle 1}{\textstyle 36} \eta - \frac{\textstyle \sqrt{5}}{\textstyle 60} \zeta & \text{for $F$=1} \\[1ex]
 \frac{\textstyle 1}{\textstyle 20} \eta + \frac{\textstyle \sqrt{5}}{\textstyle 100} \zeta & \text{for $F$=2},
 \end{cases}
\end{equation}
where
\begin{align}
\eta & =\  \mu_I^2\, \frac{20}{3}\, \frac{ \langle 5p_{3/2} \| T_1 \| 5p_{1/2} \rangle^2 }{\Delta E} \label{eta}\\
\zeta &=\ \mu_I Q\, \frac{20 \sqrt{3} }{3}\,
 \frac{ \langle 5p_{3/2} \| T_1 \| 5p_{1/2} \rangle
\langle 5p_{3/2} \| T_2 \| 5p_{1/2} \rangle }{\Delta E} . \label{zeta}
\end{align}
The magnetic dipole operator $T_1$ and electric quadrupole operator $T_2$ in the atomic reduced matrix elements
in the numerators of  Eqs.(\ref{eta}) and (\ref{zeta}) are defined in \cite[][Chap.\ V]{J:07}, and
the energy denominator $\Delta E$ in
Eqs.(\ref{eta}) and (\ref{zeta}) is the $5p_{3/2}-5p_{1/2}$ fine-structure interval.
Relativistic many-body calculations carried out in the SD approximation \cite{SJD:99} give
\begin{align*}
  \langle 5p_{3/2} \| T_1 \| 5p_{1/2} \rangle &=\ 22.3\ \text{MHz}/\mu_N\\
  \langle 5p_{3/2} \| T_2 \| 5p_{1/2} \rangle &=\ 219.6\ \text{MHz/b} .
\end{align*}
The $5p$ fine-structure interval in $^{87}$Rb  is $\Delta E$ = 7,123,020.80(5) MHz \cite{BDN:04}.
Combining these values, we find
$\eta = 3.524\ \text{kHz}$ and $\zeta = 2.916\ \text{kHz}$.

Including second-order corrections, Eqs.~(\ref{one}-\ref{three}) become
\begin{align}
\Delta W_{3} &=\ 3A + B + 8C  -\frac{1}{20}\eta -\frac{\sqrt{5}}{100}\zeta \label{bone}\\
\Delta W_{2} &=\ 2A - B - 28C +\frac{1}{45}\eta + \frac{2\sqrt{5}}{75}\zeta \label{btwo}\\
\Delta W_{1} &=\ A - B + 56C  +\frac{1}{36}\eta -\frac{\sqrt{5}}{60}\zeta  . \label{bthree}
\end{align}
Inverting these equations leads to the following second-order corrections to the previously
determined values of $A$, $B$ and $C$ listed in the second row of Table~\ref{tab1}:
\begin{align}
A^{(2)} &=\  \frac{1}{180}\eta -\frac{\sqrt{5}}{750}\zeta  = 0.0000109\ \text{MHz} \label{cone}\\
B^{(2)} &=\  \frac{1}{30}\eta +\frac{\sqrt{5}}{100}\zeta\ =0.000183\ \text{MHz} \label{ctwo}\\
C^{(2)} &=\  \frac{\sqrt{5}}{2000}\zeta\ = 0.00326\ \text{kHz} . \label{cthree}
\end{align}
The above equations for the second-order corrections are in precise
agreement with results for $^{87}$Rb presented by \citeauthor{BD:08} in Ref.~\cite{BD:08}.
The second-order corrections leave the experimental values of
$A$ and $C$ listed in the second row of Table~\ref{tab1} unchanged
and insignificantly increase the value of $B$ to 12.4944(43)~MHz.

\section{Conclusions}
In summary, the precise hyperfine measurements of the $5p_{3/2}$ hyperfine interval in $^{87}$Rb
by \citet{YSJH:96} are reanalyzed, allowing for
the possibility of a nonvanishing nuclear magnetic octupole moment. Assuming that second-order
effects in the hyperfine interaction are negligible, we obtain the values listed in the second row
for the hyperfine constants $A,\, B$ and $C$. In particular, we find a nonzero value for $C = -0.12(09)$~MHz.
Values of $A/\mu_I$, $B/Q$ and $C/\Omega$ are evaluated both empirically and using relativistic all-order
methods. The empirical calculations in combination with the experimental measurements
lead to values of $\mu_I$ and $Q$ that agree with other measured values to better than
8\%, while the all-order calculations combined with the experimental hyperfine
constants lead to values of $\mu_I$ and $Q$ that agree with other measurements
at the 2\% level.  With the aid of all-order calculations, we infer the value
$\Omega = -0.58(39)\ \mu_n$b for the nuclear magnetic octupole moment. This value (together
with the measured value of $Q$) is larger in magnitude and different in sign than the
value predicted by the nuclear shell model.  We examined the influence of the second-order hyperfine
corrections arising from the interaction between the $5p_{3/2}$ state and the neighboring $5p_{1/2}$
state. These corrections are found to make insignificant changes in the values of $A$, $B$ and
$C$ extracted from the measurements. Inasmuch as the experimental uncertainties in the values
of $C$ and $\Omega$ are relatively large, more precise measurements of the hyperfine intervals
would certainly be desirable.


\end{document}